\documentclass{optica-article}

\journal{oe}
\articletype{Research Article}
\usepackage{lineno}
\nolinenumbers
\usepackage{siunitx}
\usepackage{physics}
\usepackage{mathtools}

\begin{document}

\title{X-ray phase and dark-field computed tomography without optical elements}

\author{Thomas A. Leatham,\authormark{*} David M. Paganin,\authormark{} and Kaye S. Morgan\authormark{}}

\address{\authormark{}School of Physics and Astronomy, Monash University, Clayton, Victoria 3800, Australia}

\email{\authormark{*}thomas.leatham@monash.edu} %% email address is required; see note below about the corresponding author designation



\begin{abstract}
X-ray diffusive dark-field imaging, which allows spatially unresolved microstructure to be mapped across a sample, is an increasingly popular tool in an array of settings. Here, we present a new algorithm for phase and dark-field computed tomography based on the x-ray Fokker-Planck equation. Needing only a coherent x-ray source, sample, and detector, our propagation-based algorithm can map the sample density and dark-field/diffusion properties of the sample in 3D. Importantly, incorporating dark-field information in the density reconstruction process enables a higher spatial resolution reconstruction than possible with previous propagation-based approaches. Two sample exposures at each projection angle are sufficient for the successful reconstruction of both the sample density and dark-field Fokker-Planck diffusion coefficients. We anticipate that the proposed algorithm may be of benefit in biomedical imaging and industrial settings. 
\end{abstract}

%%%%%%%%%%%%%%%%%%%%%%%%%%  body  %%%%%%%%%%%%%%%%%%%%%%%%%%
\section{Introduction}
X-ray imaging, both in projection and in computed tomography (CT)\cite{Natterer}, is an indispensable tool for visualizing samples in a non-invasive manner. The technique has been adopted across a vast array of fields, such as medicine, security, and manufacturing. Phase-contrast x-ray imaging\cite{Endrizzi}, and its extension to computed tomography, has been a major focus for several decades within the x-ray imaging research community. In recent times, the emergence of x-ray diffusive dark-field imaging and x-ray dark-field computed tomography\cite{Pfeiffer, Wang2009,Bech2010,Prade2016,Burk2021,blykers2021} has attracted considerable interest. Diffusive dark-field imaging (`dark-field imaging' henceforth) measures a signal that arises from diffuse scattering of the illuminating x-ray wavefield from sub-pixel features within the sample, and hence allows the influence of such sub-pixel features to be detected. This approach is already finding application in clinical settings\cite{Willer2021,Gassert2023,viermetz2022}. This dark-field signal has primarily been accessed by using imaging methods that place additional optical elements in the experimental configuration, such as analyzer-based imaging\cite{Pagot2003}, grating interferometry\cite{Pfeiffer,Yashiro2010,Schaff2017}, speckle-based/structured illumination imaging\cite{MorganAnalyzer2012, Berujon2012,HWang2016,Zdora2018,Zdora2018UMPA,Alloo2022,Smith2022}, beam tracking\cite{Vittoria2015} and edge illumination\cite{Endrizzi2014,Doherty2023}. Recent work has shown that the dark-field signal may also be captured using propagation-based x-ray imaging\cite{Paganin2019, Gureyev2020,Alaleh2022,Leatham2023,ahlers2023}. 
\par
In propagation-based imaging\cite{snigirev1995,cloetens1996,Wilkins1996}, x-rays of sufficient spatial coherence are transmitted through a sample and undergo free-space propagation through a distance $\Delta$ from the sample to the detector plane. This free-space propagation converts sample-induced phase variations in the x-ray beam into intensity variations at the detector plane, which can then be measured. Additionally, unresolved microstructure contained within the volume of the sample will cause a fraction of the x-rays to be diffusely scattered, resulting in a local visibility reduction in the intensity pattern recorded at the detector, which is interpreted as dark-field signal\cite{Leatham2023}. We assume this unresolved microstructure to be spatially random, meaning that the unresolved microstructure is randomly distributed through the volume of the sample. We exclude the effects of partially ordered or fully ordered microstructure (e.g. crystals), which are beyond the scope of the present paper. Traditionally, such diffuse scattering has either not been considered or has been assumed to be negligible, in which case the image formation process in propagation-based phase-contrast imaging can be modeled by the transport-of-intensity equation (TIE)\cite{Teague1983}. Several software packages for retrieving phase information from propagation-based x-ray images based on the TIE have been developed\cite{Paganin2002, Weitkamp2011,Xtract},  which along with the adaption of TIE-based phase-retrieval algorithms to multi-material objects\cite{Beltran2010} and the use of partially spatially coherent radiation\cite{Beltran2018}, have made propagation-based imaging a widely accessible technique.
\noindent\par
\begin{figure}[htbp]
    \centering
    \includegraphics[width=0.9\linewidth]{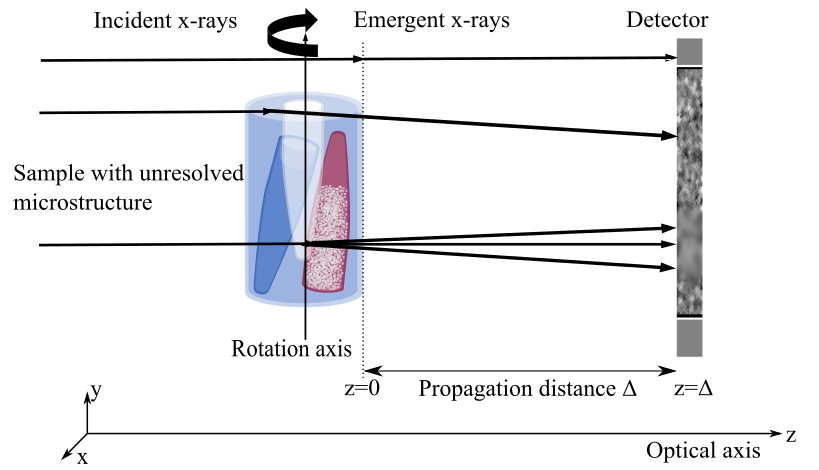}
    \caption{Experimental schematic to perform computed tomography using propagation-based imaging. The blurry region of the snippet of the image shown at the detector plane arises as a result of the sample-induced diffusion due to SAXS. This sample-induced diffusion can be extracted, along with the sample thickness, by using two intensity measurements at different propagation distances. By collecting intensity measurements at different sample rotation angles, a tomographic reconstruction of the sample density and dark-field signal can be performed.}
    \label{fig:schematic}
\end{figure}

In this paper, we extend the dark-field retrieval method presented in our previous work~\cite{Leatham2023} to CT. This approach is based on the following model for image contrast seen when a coherent x-ray wavefield propagates, namely the Fokker-Planck equation for paraxial wave optics\cite{Morgan2019,Paganin2019}:
\begin{align}
\label{eqn:FPE}
I\qty(x,y,z=\Delta)\approx I\qty(x,y,z=0)-\frac{\Delta}{k}\grad_{\perp}\vdot\qty[I\qty(x,y,z)\grad_{\perp}\phi\qty(x,y,z)]_{z=0}\nonumber\\
+\Delta^{2}\laplacian_{\perp}\qty[D\qty(x,y)I\qty(x,y,z)]_{z=0}.
\end{align}
Above, \textit{$I\qty(x,y,z=\Delta)$} is the intensity of the monochromated x-ray wavefield recorded using a detector located at a propagation distance \textit{z=$\Delta$} downstream of the sample (see Fig. \ref{fig:schematic}), \textit{$I\qty(x,y,z=0)$} is the intensity of the wavefield at the exit-surface (\textit{z=$0$}) of the sample, \textit{k} is the wavenumber of the illuminating x-ray radiation, \textit{$\phi\qty(x,y,z)$} is the phase of the x-ray wavefield, \textit{$D\qty(x,y)$} is the dimensionless diffusion coefficient describing local sample-induced small-angle x-ray scattering (SAXS), $\qty(x,y)$ denotes transverse Cartesian coordinates in planes perpendicular to the optical axis $z$ and $\grad_{\perp}\equiv (\partial/\partial x,\partial/\partial y)$ is the gradient operator with respect to $(x,y)$. In the limit of zero diffusion, i.e. $D\qty(x,y)\rightarrow 0$, the x-ray Fokker-Planck equation reduces to the TIE. Note, also, that the dimensionless diffusion coefficient may be related to the associated SAXS cone (see Fig.~1 of Ref.~\cite{Leatham2023} and Fig.~3 of Ref.~\cite{Paganin2023paraxial}) by\cite{Paganin2023paraxial}
\begin{equation}
\label{eqn:DFfactor}
D\qty(x,y)=\frac{1}{2}F\qty(x,y)\qty[\theta_{s}\qty(x,y)]^{2},
\end{equation}
where $F\qty(x,y)$ is the fraction of the x-rays that are converted to SAXS at each transverse position over the nominally planar exit-surface $z=0$ of the sample, and $\theta_{s}\qty(x,y)$ is the apex half-angle of the associated SAXS cone.
\noindent\par
The x-ray Fokker-Planck equation describes how the optical energy carried by the illuminating wavefield is transversely redistributed as it propagates downstream of the sample. A fraction of the optical energy carried by the illuminating wavefield is converted to SAXS, while the remaining fraction is coherently transported, resulting in a bifurcation into coherent and diffusive energy-flow channels. The coherent energy-flow channel is associated with phase effects (propagation-based phase-contrast), while the diffusive energy-flow channel is associated with x-ray dark-field effects (position-dependent visibility reduction). Returning to a point made earlier, phase effects are linked to the refraction of the emergent x-rays due to the sample, which results in bright/dark fringes in the recorded intensity pattern, which increase in width and visibility with propagation\cite{snigirev1995,cloetens1996}. Dark-field effects arise due to the diffuse scattering of the emergent x-rays by the unresolved spatially-random microstructure contained in the sample, with this diffuse scattering blurring the local contrast\cite{Nesterets2008}. These phase and dark-field effects are seen differently and evolve differently with propagation distance, as evidenced by the different powers of $\Delta$ that appear in the final two terms of Eq.~(\ref{eqn:FPE}). This observation is central to the reconstruction method presented in this paper. Such phase and dark-field effects can be seen in Fig.\ref{fig:schematic} below, which shows the propagation-based imaging setup that corresponds to Eq.~(\ref{eqn:FPE}). It should be noted that we use the terms `sample density' and `linear attenuation coefficient' interchangeably, where `sample density' is taken to mean `sample number density'. This equivalence is justified via the relationship (cf.~Eq.~(6) in Ref.~\cite{PaganinNeutron}):
\begin{equation}
\label{eqn:mutodensity}
\mu\qty(x,y,z)=\sigma\rho\qty(x,y,z),
\end{equation}
where $\mu\qty(x,y,z)$ is the linear attenuation coefficient, $\sigma$ is the total x-ray cross section for one scatterer and $\rho\qty(x,y,z)$ is the number density of scatterers in the sample. 
\noindent\par
As additional context for this work, we note that the concept of spatial resolution and how it can be improved is a key factor in determining the usefulness of a CT reconstruction technique. In order to improve spatial resolution, CT reconstructions that utilize the TIE phase retrieval method of Paganin et al.\cite{Paganin2002} have sometimes incorporated deconvolution filters\cite{Weitkamp2011} or taken source-size blurring into account \cite{Beltran2018}, and recently have shown an improvement in spatial resolution by incorporating the discrete mathematics used in processing digital images in the phase retrieval algorithms\cite{GPM2020, Pollock2022}. The present paper extends the method presented in Leatham et al.~\cite{Leatham2023} to the realm of CT, and also aims to show that 1) a dark-field CT can be captured without additional optics, using only propagation-based imaging, through the introduction of a tomographic Fokker-Planck linear diffusion coefficient which quantifies the diffusion of the x-ray beam by a given voxel of the sample, and 2) by separating out phase and dark-field effects, an improvement in spatial resolution can be seen in the recovered sample density image when compared to the TIE-based phase retrieval method of Paganin et al.~\cite{Paganin2002}. Such an improvement in spatial resolution may be anticipated for the following reason. Since the projected intensity pattern is locally blurred by dark-field effects, the spatial resolution of both the sample thickness projections and CT sample density slices is reduced. If dark-field effects are not taken into account, such as in the TIE phase retrieval method~\cite{Paganin2002}, the spatial resolution of the recovered images suffers.  
\noindent \par
The paper is structured as follows. Section 2 details our Fokker-Planck reconstruction method for x-ray phase and dark-field CT without optical elements, and section 3 shows the results of the experiment we performed to demonstrate the ability of our method to tomographically reconstruct phase and dark-field signals in three dimensions. Section 4 discusses the broader implications of our work and section 5 outlines avenues for future research, as well as providing some concluding remarks.  
\section{Method}
\label{sec:Method}
The CT reconstruction method we propose here builds directly on the phase and dark-field retrieval method presented in Leatham et al.~\cite{Leatham2023}. In this approach, a single-material approximation is used to write the sample-exit-surface intensity and phase terms in the Fokker-Planck equation as a function of sample thickness only. Two unknowns, the sample projected thickness $T$ and sample diffusion coefficient $D$, are then determined by solving this single-material Fokker-Planck equation using data obtained at two different propagation distances $\Delta_{1}$ and $\Delta_{2}$. Given such propagation-based intensity images captured at two different sample-to-detector propagation distances ($\Delta_{1}\neq \Delta_{2}$), at a fixed energy, over a set of sample projection angles, denoted by $\Theta$, and assuming quasi-monochromatic $z$-directed plane waves of incident intensity $I_{0}$ illuminate the sample, our method for dark-field CT constitutes the following. Firstly, using the two propagation-based intensity images captured at each projection angle, we calculate the sample transmission at each angle, $t\qty(x,y;\Theta)$, according to (cf. Ref.~\cite{Leatham2023}):
\begin{align}
\label{eqn:thickness}
t\qty(x,y;\Theta)=\exp\qty(-\int \mu\qty(x,y,z)dz)=\mathcal{F}^{-1}\!\qty[\frac{\mathcal{F}\qty[\Delta_{2}^{2}I\qty(x,y,z=\Delta_{1};\Theta)-\Delta_{1}^{2}I\qty(x,y,z=\Delta_{2};\Theta)]}{I_0\qty(\Delta_{2}^{2}-\Delta_{1}^{2}+\frac{\gamma}{2k}\Delta_{1}\Delta_{2}\qty(\Delta_{2}-\Delta_{1})\qty(k_{x}^{2}+k_{y}^{2}))\!}],
\end{align}
where $\gamma=\delta\qty(x,y,z)/\beta\qty(x,y,z)$= \textit{const}, $\delta\qty(x,y,z)$ is the real decrement of the sample complex refractive index, $n\qty(x,y,z)=1-\delta\qty(x,y,z)+i\beta\qty(x,y,z)$, and $k$ is the wavenumber corresponding to a given energy of x-ray photons. The value for $\gamma$  is chosen for `single-material' phase retrieval, however any additional material boundaries within the sample will simply be either slightly over or under-blurred. Note that for a single-material sample, the sample transmission can be written as $t\qty(x,y;\Theta)=\exp\qty[-\mu_{s}T\qty(x,y;\Theta)]$, where $\mu_{s}=2k\beta$ is the linear attenuation coefficient of the sample. This expression can be compared to our earlier work deriving Fokker-Planck thickness retrieval for projection images\cite{Leatham2023}. In that paper, the single-material sample transmission expression given above was effectively substituted into Eq.~(\ref{eqn:thickness}) to solve for the sample projected thickness. While it is was not necessary to calculate the sample projected thickness as a part of tomographic reconstruction process, we used the single-material sample transmission expression to generate the thickness profiles seen in Fig.~\ref{fig:Projection}(g). These thickness profiles are more closely related to the observed sample, and make comparison with conventional TIE-based sample thickness retrieval\cite{Paganin2002} easier. While the value of $\mu_{s}$ is important in determining the sample projected thickness, this quantity is effectively canceled out when computing the sample transmission and as such does not impact the distribution of the linear attenuation coefficient, $\mu\qty(x,y,z)$, reconstructed from CT. Secondly, from the calculated sample transmission using Eq.~(\ref{eqn:thickness}) above, along with one of the captured propagation-based intensity images, we compute the dark-field signal at each projection angle via (cf. Ref~\cite{Leatham2023}):
\begin{align}
\label{eqn:DF}
D&\qty(x,y;\Theta)=t^{-1}\qty(x,y;\Theta)\grad_{\perp}^{-2} \qty[\frac{I\qty(x,y,z=\Delta_{1};\Theta)}{I_{0}\Delta_{1}^{2}}
-\qty(\frac{1}{\Delta_{1}^{2}}-\frac{\gamma}{2k \Delta_{1}}\laplacian_{\perp})t\qty(x,y;\Theta)].
\end{align} Note that the inverse Laplacian operator in Eq.~(\ref{eqn:DF}) is defined via
\begin{equation}
\label{eqn:invlap}
\grad_{\perp}^{-2}=-\mathcal{F}^{-1}\frac{1}{k_{x}^{2}+k_{y}^{2}}\mathcal{F},
\end{equation}
with the replacement 
\begin{equation}
\label{eqn:replace}
\frac{1}{k_{x}^{2}+k_{y}^{2}}\rightarrow \frac{1}{k_{x}^{2}+k_{y}^{2}+\varepsilon}
\end{equation}
used in practice to avoid the singularity at $\qty(k_{x},k_{y})=\qty(0,0)$. The operator \textit{$\mathcal{F}$} denotes a Fourier transformation  with respect to \textit{$x$} and \textit{$y$}, with the corresponding Fourier-space coordinates being denoted by \textit{$(k_x,k_y)$}, while \textit{$\mathcal{F}^{-1}$} denotes inverse Fourier transformation with respect to \textit{$k_x$} and \textit{$k_y$}. Here, \textit{$\varepsilon > 0$} is small compared to \textit{$k_{x}^{2}+k_{y}^{2}$}, except in the vicinity of the origin of Fourier space. We use the Fourier transform convention found in Refs.~\cite{Paganin2002} and \cite{Coherent}. 
\noindent \par
\begin{figure}[htbp]
    \centering
    \includegraphics[width=\linewidth]{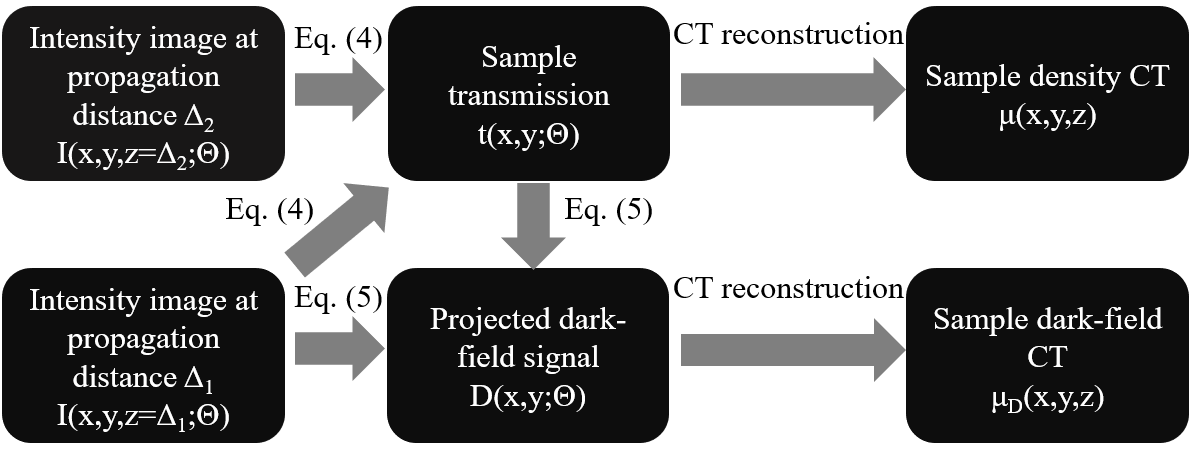}
    \caption{Flowchart summarizing the propagation-based method described in this section for dark-field and phase CT.}
    \label{fig:flow_overview}
\end{figure}

The final step of our algorithm is to perform CT reconstruction from the calculated sample transmission and dark-field signals. In particular, we input the calculated sample transmissions into the CT reconstruction code of Ref.~\cite{Xtract} to obtain the three-dimensional distribution of the linear attenuation coefficient, $\mu\qty(x,y,z)$, for instance by using filtered-back-projection\cite{KakSlaney}. The parameter $\gamma$ determines the sharpness of boundaries between adjacent materials in the reconstructed volume. In order to perform CT reconstruction from the projected dark-field signals, we define the quantity $\mu_{D}$, which we refer to as the `Fokker-Planck linear diffusion coefficient', via
\begin{equation}
\label{eqn:linDFcoeffCT}
D\qty(x,y;\Theta)=\int \mu_{D}\qty(x,y,z) dz.
\end{equation}
The quantity $\mu_{D}$ quantifies how much a given voxel of the sample diffuses the x-ray beam, independent of the set-up used to capture this measurement. From the projected dark-field signals given by Eq.~(\ref{eqn:DF}), we create projections of $\exp\qty[-D\qty(x,y;\Theta)]$, which we input into the CT reconstruction code of Ref.~\cite{Xtract} in order to obtain the three-dimensional distribution of the Fokker-Planck linear diffusion coefficient, $\mu_{D}$, on account of Eq.~(\ref{eqn:linDFcoeffCT}). Note that the CT code used in this paper, and indeed most CT codes, assume that the input is `transmitted intensity', and hence take the negative logarithm of the inputted projections. Hence, having exponentiated images, such as sample transmission given by Eq. (4) and projections of $\exp\qty[-D\qty(x,y;\Theta)]$, is crucial to obtaining the desired distribution of $\mu\qty(x,y,z)$ and $\mu_{D}\qty(x,y,z)$ from such CT codes. A summary of our algorithm workflow can be found in Fig.~\ref{fig:flow_overview}.
\noindent \par
In the case of a single-material sample with position-dependent projected thickness $T$, whose statistical properties are independent of depth $z$ for any particular transverse location $\qty(x,y)$, Eq.~(\ref{eqn:linDFcoeffCT}) reduces to
\begin{equation}
\label{eqn:linDFcoeff}
D=\mu_{D}T,
\end{equation}
where functional dependencies have been omitted. The result given by Eq.~(\ref{eqn:linDFcoeff}) could be anticipated based on the following logic. The root-mean-square (RMS) apex half-angle, $\theta_{s}$ is proportional to the square root of the sample thickness\cite{Paganin2023paraxial}. Given the relationship in Eq.~(\ref{eqn:DFfactor}), this implies that the dark-field signal is proportional to the sample thickness. Taking the Fokker-Planck linear diffusion coefficient, $\mu_{D}$, to be the factor of proportionality, one arrives at Eq.~(\ref{eqn:linDFcoeff}). 
\noindent \par
To close this section, we link the Fokker-Planck linear diffusion coefficient to the dark-field extinction coefficient, as employed in grating interferometry. Imaging methods that make use of reference patterns, for example, grating-based imaging or single-grid imaging, often speak in terms of a visibility reduction that arises due to unresolved sample microstructure contained in the sample\cite{Bech2010, Lynch2011}. This visibility reduction, $V$, can be expressed in terms of the dark-field signal via (see Eq.~(132) of Ref.~\cite{Paganin2023paraxial}, also mentioned in Ref.~\cite{Morgan2019})
\begin{equation}
\label{eqn:dftoV}
V=\exp\qty(-\frac{4\pi^{2}D\Delta^{2}}{p^{2}}),
\end{equation}
where $p$ is the period of the analyzer grating. This visibility reduction is used to tomographically quantify grating-interferometry-based dark-field via the broadly-adopted sample linear diffusion coefficient / dark-field extinction coefficient, $\mu_{d}$, for a single-material sample\cite{Bech2010,Lynch2011}:
\begin{equation}
\label{eqn:Vtomud}
V=\exp\qty(-\mu_{d}T).
\end{equation}
Equating Eq.~(\ref{eqn:dftoV}) and Eq.~(\ref{eqn:Vtomud}), and solving for $\mu_d$, one obtains:
\begin{equation}
\label{eqn:mudtoD}
\mu_{d}=\frac{4\pi^{2}\Delta^{2}D}{p^{2}T}=\frac{4\pi^{2}\Delta^{2}\mu_{D}}{p^{2}},
\end{equation}
where Eq.~(\ref{eqn:linDFcoeff}) has been used in the second step. As the propagation distance, $\Delta$, and the period of the reference pattern, $p$, are properties of the experimental setup, Eq.~(\ref{eqn:mudtoD}) provides a way to convert between the Fokker-Planck linear diffusion coefficient $\mu_{D}$, which is independent of experimental setup parameters, to the more broadly-adopted set-up-dependent linear diffusion coefficient, $\mu_{d}$. 
\section{Experimental results}
\begin{figure}[htbp]
    \centering
    \includegraphics[width=\linewidth]{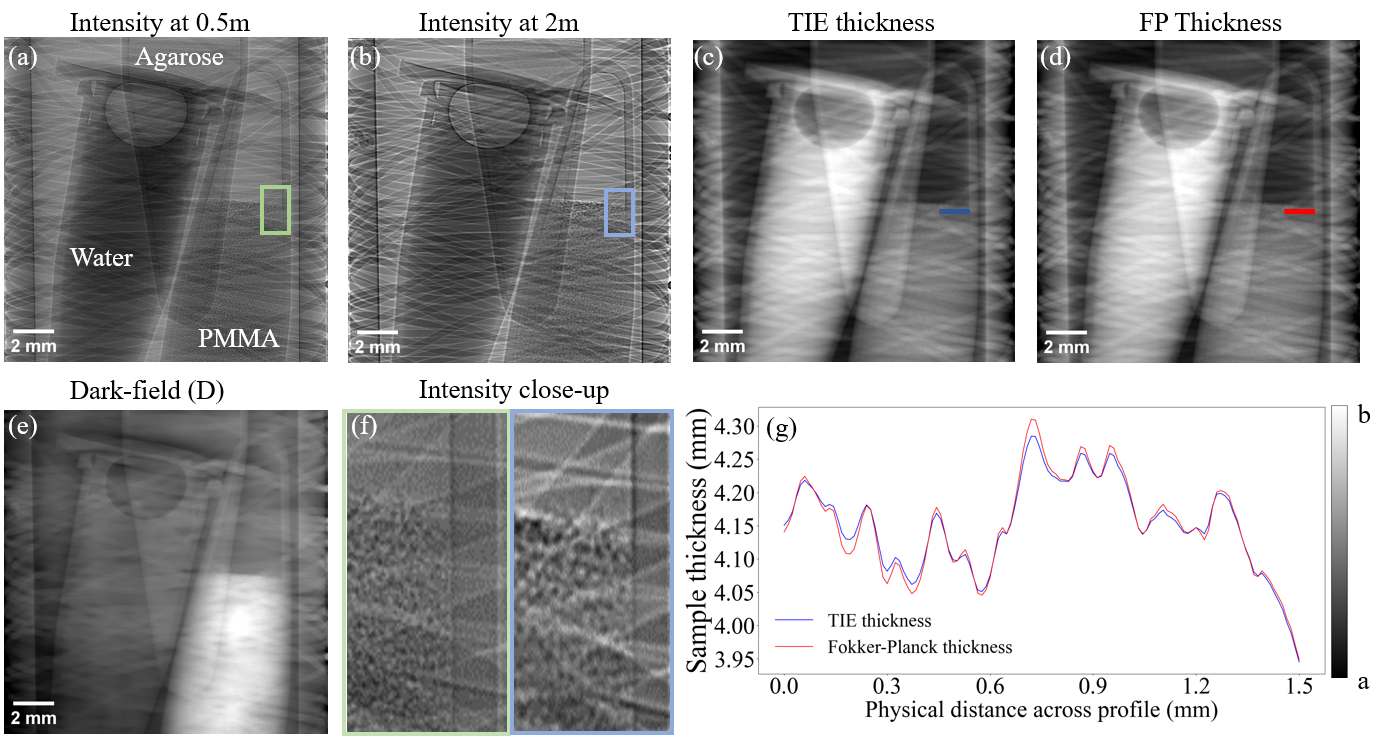}
    \caption{Experimental images of the sample, which includes three PMMA tubes, containing water, agarose powder and microspheres, from left to right, shown here in projection, taken at $\Theta=0^\circ$. All images are shown on a linear grayscale, where $a$ is the minimum value and $b$ is the maximum value. Top row: propagation-based intensity images taken at (a) \SI{0.5}{\meter} propagation distance and (b) \SI{2}{\meter} propagation distance (for the grayscale $a=0.6$, $b=1.1$), and sample thickness reconstructions using (c) TIE phase retrieval from \SI{0.5}{\meter} intensity data and (d)  Fokker-Planck (FP) phase retrieval via Eq.~(\ref{eqn:thickness}) ($a=$ 2 mm, $b=$ 8 mm). Bottom row: (e) dark-field ($D$) reconstruction using the projection image captured at a propagation distance of \SI{2}{\meter} ($a=0$, $b=7.07\times 10^{-11}$), (f) close-up regions of the green and blue boxes in (a) and (b), respectively and (g) sample thickness profiles taken over the blue and red lines in panels (c) and (d).}
    \label{fig:Projection}
\end{figure}

In order to demonstrate the applicability of our CT reconstruction method outlined in the previous section, we collected propagation-based x-ray phase-contrast images at the Australian Synchrotron, on the Imaging and Medical Beamline (IMBL) in Hutch 3B. The sample consisted of a \SI{1.3}{\centi\meter} diameter Polymethyl Methacrylate (PMMA) tube with nylon wire wrapped around the outside of the tube, creating some background texture that was part of the sample, as well as three PMMA microtubes sitting inside the larger tube. With reference to the $0^\circ$ projections (see Fig.~\ref{fig:Projection}), the bottom-left microtube contained water, the bottom-right microtube contained PMMA microspheres and the upper microtube contained agarose powder. The PMMA microspheres were supplied from the company \textit{Polysciences Inc.} (catalogue number 26305-500). The diameters of the microspheres were distributed within the range 1-\SI{10}{\micro\meter}, with a mean diameter of \SI{6}{\micro\meter}, and the density of the microspheres was approximately $\SI{1.19}{\gram/\centi\meter^{3}}$. The agarose powder contained particles spanning a relatively large range of diameters, with these particles being heterogeneous in size and shape. The diameters of the particles contained in the agarose powder were distributed between 20-\SI{160}{\micro\meter}, with an average diameter of approximately \SI{80}{\micro\meter}. The sample was placed on a dedicated table located approximately \SI{130}{\meter} from the source, where x-ray photons of energy \SI{25}{\kilo\electronvolt} were produced from a \SI{2}{\tesla} dipole bending magnet. At this energy, the wavenumber is $k=\SI{1.27e-11}{\meter^{-1}}$ and the complex refractive index of PMMA has $\delta=\num{4.26e-07}$ and $\beta=\num{1.81e-10}$, so that $\gamma=2353.96$ and $\mu_{s}=\SI{45.9}{\meter^{-1}}$. This value of $\gamma$ was used to control the sharpness of the boundaries that appear in our sample when calculating the sample transmission, and subsequently the sample projected thickness, as well as the projected dark-field signals. 
\noindent \par
The IMBL's `Ruby' detector\cite{Ruby2013}, used to image the sample, is equipped with a commercial macro lens with a variable focal distance, allowing for the pixel size to be set between \SI{6.3}{\micro\meter} and \SI{22.3}{\micro\meter}. For our experiment, the pixel size used was \SI{12.3}{\micro\meter}. We moved the detector to distances of \SI{0.5}{\meter} and \SI{2}{\meter} downstream of the sample, and performed a scan of the sample at each propagation distance, with each scan containing $1810$ projections collected over $181^\circ$, with a rotation step of $0.1^\circ$ between each projection. These propagation distances provided a balanced trade-off between visualizing phase and dark-field effects, with the phase effects visually apparent at both distances and subtle dark-field blurring effects seen at the larger distance. We also collected $30$ images of both the flat-field and the dark-current during each scan. Each scan took approximately $15$ minutes to complete, noting that we made no particular effort to minimize radiation dose in this first proof-of-principle experiment.
\noindent \par
Prior to analysis, the projections captured at each distance were flat-field and dark-current corrected, then resized to account for slight magnification differences using the source-to-sample distance and finally registered to sub-pixel accuracy using the `phase\_cross\_correlation' function of the open-source python library scikit-image\cite{scikit}, with an upsampling factor of $1000$ used as the argument of this function. All image processing and data analysis was performed using Python3 code and X-tract software\cite{Xtract} on the Australian Synchrotron Compute Infrastructure (ASCI) desktop. Taking each pair of propagation-based images (Figs. \ref{fig:Projection}(a)-(b)), Eqs.~(\ref{eqn:thickness}) and (\ref{eqn:DF}) were used to reconstruct two-dimensional projection images of the sample transmission (and hence single-material thickness) (Fig.~\ref{fig:Projection}(d)) and dark-field signal (Fig.~\ref{fig:Projection}(e)) at each projection angle. The sample thickness reconstruction calculated using the TIE phase retrieval\cite{Paganin2002} is also provided in Fig.~\ref{fig:Projection}(c) for comparison.  
\noindent\par
Panel (f) of Fig.~\ref{fig:Projection} demonstrates the increase in strength of dark-field effects seen with increasing propagation distance, which causes a local visibility reduction in the intensity image captured at \SI{2}{\meter}, as compared to the intensity image captured at \SI{0.5}{\meter}\cite{Leatham2023} . This local visibility reduction is seen as regions of increased blurriness in the measured intensity data, with the edges of the wrapped fibres becoming less visible, in particular where the projected thickness of microstructures increases (e.g. bottom-left of these regions relative to top-right). Figure \ref{fig:Projection}(c)/(d) shows that the sample thickness reconstructions constructed using TIE phase retrieval\cite{Paganin2002} and our present method appear to be identical by eye. However, as discussed in Leatham et al.\cite{Leatham2023}, a Fokker-Planck approach to thickness reconstruction provides increased spatial resolution relative to TIE phase retrieval, since TIE phase retrieval misinterprets diffusive dark-field contrast as slowly changing sample thickness. This can be seen from panel (g), where the blue density profile (TIE) and the red density profile (Fokker-Planck) are very similar, but the blue/TIE profile is more slowly-varying on the left of the line profile where dark-field signal coming from the PMMA microspheres locally reduces contrast/resolution. The Fokker-Planck-based reconstruction of the sample transmission using Eq.~(\ref{eqn:thickness}) at each rotation angle serves a dual purpose. Firstly, the computed set of sample transmissions can be input into a computed tomography reconstruction to create a map of the linear attenuation coefficient. Secondly, the sample transmission at each rotation angle is used as a part of the calculation of the dark-field signal at each rotation angle according to Eq.~(\ref{eqn:DF}). 
\noindent\par 
Using the reconstructed sample transmission projections and the measured intensity data at \SI{2}{\meter}, we computed the dark-field signal projections arising from SAXS, induced by the unresolved sample microstructure according to Eq.~(\ref{eqn:DF}). It is evident from Fig.~\ref{fig:Projection}(e) that the PMMA microspheres generate the strongest dark-field signal, with a weaker diffuse scattering signal coming from the agarose powder and the water. We suspect that the total volume of agarose powder is not sufficient to provide a very strong dark-field signal, based on the weak contrast observed in the projected thickness image shown in Fig.~\ref{fig:Projection}(d), corresponding to a weak attenuation signal. Due to the heterogeneity in the shape of the particles that made up the agarose powder, the agarose powder likely did not pack as efficiently as the PMMA microspheres, meaning there were not as many air/powder interfaces as seen for the PMMA microspheres, resulting in this weaker dark-field signal. Additionally, with particle diameter ranging from 20-\SI{160}{\micro\meter}, fewer particles would be seen in a given projection than for the 1-\SI{10}{\micro\meter} PMMA microspheres. The weak dark-field signal emanating from the agarose powder may be better retrieved by using a larger propagation distance in the dark-field reconstruction. By comparison, it can be seen that the projected sample thickness is largest in the microtube containing water, from Fig.~\ref{fig:Projection}(d). This demonstrates the complementary nature of the sample thickness and the dark-field. Note that some fully-resolved features appear in the projected dark-field signal shown in Fig.~\ref{fig:Projection}. The appearance of such fully-resolved features may be due to slight mismatches in the fringes of the measured intensity images (e.g. Fig.~\ref{fig:Projection}(a) and (b)) and the `no dark-field' estimated intensity image given by the $\qty(\frac{1}{\Delta_{1}^{2}}-\frac{\gamma}{2k \Delta_{1}}\laplacian_{\perp})t\qty(x,y;\Theta)$ term in Eq.~(\ref{eqn:DF}). Additionally, such fully-resolved features may provide a weak edge-scattering signal, which can contribute to the projected dark-field signal here, since edge-scattering is not explicitly taken into account in our present analysis. Fully-resolved features, and any background signal more generally, could potentially be removed from the projected dark-field signal by including in the field of view a region which is known to not scatter x-rays, or regions where the sample is absent. Such regions can act as a reference point of zero dark-field when reconstructing the dark-field in projection, which could be utilized by subtracting the mean value of the reconstructed dark-field signal in such regions\cite{Paganin2023paraxial}. Note that since the dark-field signal is retrieved by inverting the Laplacian operator, the point of zero dark-field is ambiguous without such a potential reference point. 
\noindent\par
\begin{figure} [htbp]
    \centering
    \includegraphics[width=\linewidth]{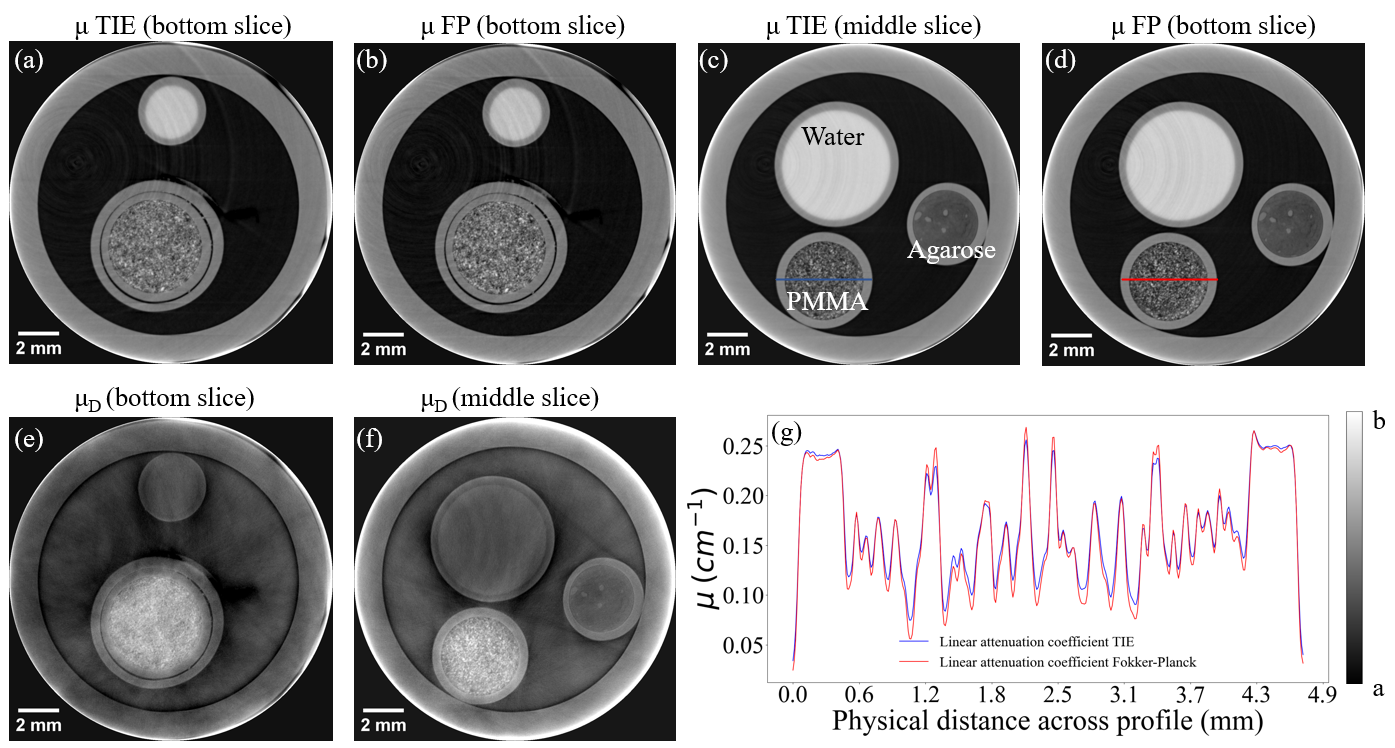}
    \caption{CT reconstructions of our sample. Top row: slices of the linear attenuation coefficient reconstructed using  (a) TIE phase-retrieval (bottom slice), (b) Fokker-Planck (FP) phase-retrieval Eq.~(\ref{eqn:thickness}) (bottom slice), (c) TIE phase-retrieval (middle slice) and (d) FP phase-retrieval Eq.~(\ref{eqn:thickness}) (middle slice). The grayscales are shown with $a=\SI{0}{\centi\meter^{-1}}$, $b=\SI{0.48}{\centi\meter^{-1}}$ for (a)-(d). Bottom row: (e) bottom slice of the Fokker-Planck linear diffusion coefficient, (f) middle slice of the Fokker-Planck linear diffusion coefficient ($a=\SI{0}{\centi\meter^{-1}}$, $b=\SI{7.53e-11}{\centi\meter^{-1}}$) and (g) profiles taken over the blue and red lines in (c) and (d), showing a spatial resolution improvement in the reconstruction of the linear attenuation coefficient when using the FP method compared to TIE phase retrieval.}
    \label{fig:CT}
\end{figure}

We then used a Fast Filtered-Back-Projection based on a grid reconstruction tomography scheme\cite{Dowd1999} with a Hamming window to create maps of the linear attenuation coefficient and Fokker-Planck linear diffusion coefficient from the full set of projections, as outlined in the method section of this paper, via Eq.~(\ref{eqn:linDFcoeffCT}). Slices of the linear attenuation coefficient and Fokker-Planck linear diffusion coefficient through the bottom and middle of our sample can be found in Fig.~\ref{fig:CT}. 
\noindent\par
To the eye, the CT slices of the linear attenuation coefficient shown in Fig.~\ref{fig:CT}(a) and (b), as well as in (c) and (d), appear identical. However, as can be seen from the intensity profile given in panel (g), there is a spatial resolution improvement in the reconstruction of the linear attenuation coefficient when using the Fokker-Planck method described here compared to TIE-based phase retrieval, which comes directly from the increase in the spatial resolution of the projected sample thickness. This can also be seen by noting that one can get from the Fokker-Planck slice (Fig.~\ref{fig:CT}(d)) to the TIE slice (Fig.~\ref{fig:CT}(c)) by blurring with a two-dimensional symmetric Gaussian function with a standard deviation of 1-2 pixels. This increase in spatial resolution is ultimately a consequence of the Fokker-Planck approach taken to reconstructing the sample transmission (as in our method), and hence sample thickness, which unlike TIE phase retrieval, distinguishes between phase and dark-field effects. It is worth reiterating that the increase in spatial resolution found here is only due to separating out phase and dark-field effects using two intensity images. This is in contrast to how spatial resolution improvement in CT reconstructions from TIE phase retrieval has been achieved in the past, primarily through using deconvolution filters to qualitatively boost spatial resolution\cite{Weitkamp2011}. Such an approach could be added to further increase spatial resolution, as appropriate. The use of discrete mathematics\cite{GPM2020, Pollock2022}, through the discrete Fourier transform, may also further enhance the spatial resolution increase seen in our Fokker-Planck approach. The other key result that arises from explicitly taking dark-field effects into account is the ability of our method to tomographically reconstruct a dark-field signal in three dimensions, through the Fokker-Planck linear diffusion coefficient. The reconstructed slices of the Fokker-Planck linear diffusion coefficient, Fig.~\ref{fig:CT}(e) and Fig.~\ref{fig:CT}(f), highlight a strong diffusive signal from the PMMA microspheres through both the top and the middle of the sample, while the water and the agarose powder provide a much weaker signal. The diffusive signal seen across the tube containing the PMMA microspheres is relatively uniform, which reflects the relatively uniform packing of the microspheres. The Fokker-Planck linear diffusion coefficient provides  information about the sample which is complementary to the linear attenuation coefficient, much in the same way that the projected dark-field signal is complementary to the projected sample thickness. The non-uniform background seen within the sample holder may be due to slight mismatches in the high-contrast fringes within the measured intensity images and the `no dark-field' estimated intensity image in projection, which are then back-projected. Additionally, image noise may also contribute to this non-uniform background, as image noise can induce a dark-field signal\cite{Paganin2023paraxial}. The apparent dark-field signal seen from the PMMA microtubes is likely an edge signal. This apparent dark-field signal may be due to the absence of a reference point of zero dark-field, as well as slight mismatches in the intensity fringes (described above), across the PMMA microtubes. 
\section{Discussion}
This paper presents a new CT phase and dark-field reconstruction algorithm for two-distance propagation-based imaging, which builds upon the Fokker-Planck-based retrieval method of Leatham et al.\cite{Leatham2023}. To our knowledge, the use of propagation-based imaging for diffusive dark-field computed tomography has only been demonstrated in papers by Gureyev et al. \cite{Gureyev2020} and Aminzadeh et al. \cite{Alaleh2022}. A comparison between the method presented in this manuscript and that presented in Refs.~\cite{Gureyev2020} and \cite{Alaleh2022} would be an interesting pursuit for further investigation. Of interest in future work would also be a comparison of the method presented here to dark-field CT methods which make use of optical elements, such as those methods utilizing grating interferometry\cite{GICT} or analyzer-based imaging\cite{MIR2006}. As some preliminary remarks on this front, we note that dark-field CT methods employing additional optical elements in the experimental setup have traditionally required multiple sample exposures, in the vicinity of seven exposures, associated with a higher sample dose (noting that a sliding window approach \cite{Zanette2013} can help  with this). Our method, on the other hand, only needs two sample exposures for each tomographic projection, and hence may provide a dose-saving advantage over such dark-field CT methods. The impact on spatial resolution would have to be investigated. Additionally, with regards to spatial coherence requirements, our method requires that the characteristic length scale associated with source-size blurring be less than the characteristic length scale associated with diffuse scattering due to sample-induced SAXS, so that the diffusive dark-field signal associated with position-dependent SAXS can be detected\cite{Leatham2023}. 
\noindent \par
In order to transition the method presented in this manuscript to clinical or industrial settings, potential optimization conditions would need to be addressed. Perhaps the biggest limiting factor of our method is the need for two sample exposures at each rotation angle, as this affects the total scan time. Ideally in the clinical and industrial settings, one would like to minimize sample dose, and for the scan to be as quick as possible, so that a large number of samples can be examined in a given amount of time. One way in which the need for two sample exposures could be removed in our method is by utilizing an imaging setup with either a beam-splitting geometry or a semi-transparent detector\cite{Carnibella2012}. Two detectors placed at the desired propagation distances would be required in both geometries, with an object capable of reflecting and transmitting the x-rays being inserted after the first detector in the beam-splitting geometry. The first detector would need to be semi-transparent in the semi-transparent detector geometry, so as to allow ideally 50\% of the x-rays to pass through to the second detector. In each geometry, the two detectors would be fixed at the chosen propagation distances and so in principle there should be no image artifacts arising from misalignment between the two propagation distances. Each of these imaging geometries would allow for propagation-based intensity images to be captured at two propagation distances at the same time, i.e. with a single sample exposure. These imaging geometries may also allow our method to be integrated with previous work demonstrating how a hundred or even a thousand CT projections could be captured per second in phase-contrast computed tomography\cite{Foamingmetal2019,1000ct}, thereby enabling time-resolved tomography studies. 
Fast dark-field tomography could alternatively be achieved by applying the $\mu_D$ approach described here to a recently-published variation on Fokker-Planck propagation-based dark-field imaging, where two energies are used instead of two propagation distances \cite{ahlers2023}. %It is worth noting that while neither a beam-splitting geometry nor a semi-transparent detector setup\cite{Carnibella2012} has been implemented in a propagation-based dark-field imaging setup to date, it is possible that such technology may exist within the immediate future, given the historical rapid expansion of the x-ray imaging field. 
\noindent\par
Propagation-based imaging and in particular, the TIE phase retrieval method of Paganin et al.\cite{Paganin2002}, has been applied not only in the context of x-ray imaging, but also to visible-light microscopy\cite{Poola2017}, electron microscopy\cite{Liu2011} and neutron imaging\cite{PaganinNeutron}. Given that the phase and dark-field retrieval method of Leatham et al.\cite{Leatham2023}, which underpins the method presented in this paper, is a diffusive generalization of the method of Paganin et al.~and makes use of the same imaging setup, it stands to reason that the method of Leatham et al.~and by extension the method contained in this paper, may also have a domain of validity that extends beyond x-rays and into visible-light, electron and neutron tomography.
\section{Directions for future research}
Here we outline several possible directions for future research based on the work presented in this paper. The first of these potential directions is to generalize our CT reconstruction method to samples composed of arbitrarily many materials. For such samples, the key quantities to be reconstructed are the attenuation of the illuminating wavefield, the phase shift of the illuminating wavefield induced by the sample, and the dark-field signal. This could be achieved by using three different propagation distances to disentangle the attenuation, phase and dark-field signals in Eq.~(\ref{eqn:FPE}) in projection. One could then use standard CT techniques to create maps of the sample linear attenuation coefficient, $\mu\qty(x,y,z)$, the real decrement of the complex refractive index of the sample, $\delta\qty(x,y,z)$, and the Fokker-Planck linear diffusion coefficient, $\mu_D\qty(x,y,z)$.
\noindent\par
A second possible direction would be to extend the CT reconstruction method presented in this paper to the case of diffusion tensor tomography by considering samples which contain sub-pixel features that scatter in a preferred direction, i.e. samples that produce a `directional dark-field' signal\cite{JensenMay2010,JensenDec2010}. This could be achieved by replacing the single diffusion coefficient $D(x,y)$ with a dimensionless rank-two symmetric diffusion tensor\cite{Paganin2019,Paganin2023paraxial}:
\begin{equation}
\label{eqn:Dtensor}
\small
D\qty(x,y)\rightarrow 
\Tilde{\vb{D}}\qty(x,y)=\mqty[\Tilde{D}_{xx}\qty(x,y) & \Tilde{D}_{xy}\qty(x,y) \\ \Tilde{D}_{xy}\qty(x,y) & \Tilde{D}_{yy}\qty(x,y)].
\end{equation}
This modifies Eq.~(\ref{eqn:FPE}) to\cite{Morgan2019,Paganin2019,Paganin2023paraxial}:
\begin{align}
\label{eqn:directionalFPE}
I\qty(x,y,z=\Delta)=I\qty(x,y,z=0) &-\frac{\Delta}{k}\grad_{\perp}\vdot\qty[I\qty(x,y,z)\grad_{\perp}\phi\qty(x,y,z)]_{z=0} \nonumber\\
&+\Delta^{2}\grad_{\perp}\vdot\qty[\Tilde{\vb{D}}\qty(x,y)\grad_{\perp}I\qty(x,y,z)]_{z=0}.
\end{align}
Here, the transverse gradient $\grad_{\perp}$ is to be understood as the column-vector operator $\qty(\pdv{}{x},\pdv{}{y})^T$, where a superscript $T$ denotes matrix transposition, while the transverse divergence $\grad_{\perp}\vdot$ should be understood as the row-vector operator $\qty(\pdv{}{x},\pdv{}{y})$. To perform CT reconstruction from Eq.~(\ref{eqn:directionalFPE}), one would first need to solve Eq.~(\ref{eqn:directionalFPE}) for the diffusion tensor and the sample phase in projection, and then create maps of the sample linear attenuation coefficient, the real decrement of the sample complex refractive index and Fokker-Planck linear diffusion coefficient corresponding to each component of the diffusion tensor through standard CT techniques. 
\noindent\par
Additionally, the method presented in this paper might be applied to lung imaging. Recent studies, such as that by Willer et al.\cite{Willer2021}, have demonstrated that dark-field chest x-rays have the potential to differentiate between healthy and diseased lung tissue in humans, while conventional chest x-rays cannot. Propagation-based methods are well-suited to smaller samples, with a number of small-animal biomedical research studies using a propagation-based set-up to better understand lung health\cite{Siew2009,Stahr2016,Gradl2018}. For this reason, it would be an interesting avenue for future research to investigate the ability of our CT method to (i) differentiate between healthy and diseased lung tissue using animal subjects for a variety of different lung disease models, and (ii) to quantify the extent of the disease. Our method could also be used in biomedical research, for example to measure the effectiveness of treatments for lung diseases\cite{MorganCF2014}, or to assess the severity of disease\cite{Frank2022}.
\noindent\par
A further potential interesting avenue for future research would be to investigate how the three-dimensional map of the Fokker-Planck linear diffusion coefficient could be used to quantify properties of the sample microstructure. Such properties could include the size or surface-area-to-volume ratio of the microstructure, $a\qty(x,y,z)$\cite{Paganin2023paraxial}. A reconstruction of such properties can be seen as a `removal' of the influence of the x-ray beam, as the Fokker-Planck linear diffusion coefficient encodes both the influence of the x-ray beam used to illuminate the sample and the properties of the sample, while the surface-area-to-volume ratio, for example, is purely a sample property. Since microstructure size may also be determined from the sample linear diffusion coefficient ($\mu_{d}$) in grating-based imaging\cite{Lynch2011}, it is natural to wonder whether there is a correspondence between propagation-based CT as described in this paper and imaging methods which use reference patterns. Indeed, since dark-field signal in the former context and visibility reduction in the latter context are related via Eq.~(\ref{eqn:dftoV}), and $\mu_{d}$ can be obtained via\cite{Bech2010}
\begin{equation}
V\qty(x,y;\Theta)=\exp(-\int \mu_{d}\qty(x,y,z)dz),
\end{equation}
we have the correspondence as shown in Fig.~\ref{fig:SAVR_from_df} below. Note that converting a dark-field signal to a visibility reduction can be seen as `adding in' the properties of the reference pattern/mask to the x-ray beam and sample properties, while the process of inferring sample properties from $\mu_{d}$ can be seen as a `removal' of the influence of the x-ray beam and reference pattern/mask properties, leaving only the sample properties. 
\begin{figure}[htbp]
    \centering
    \includegraphics[width=\linewidth]{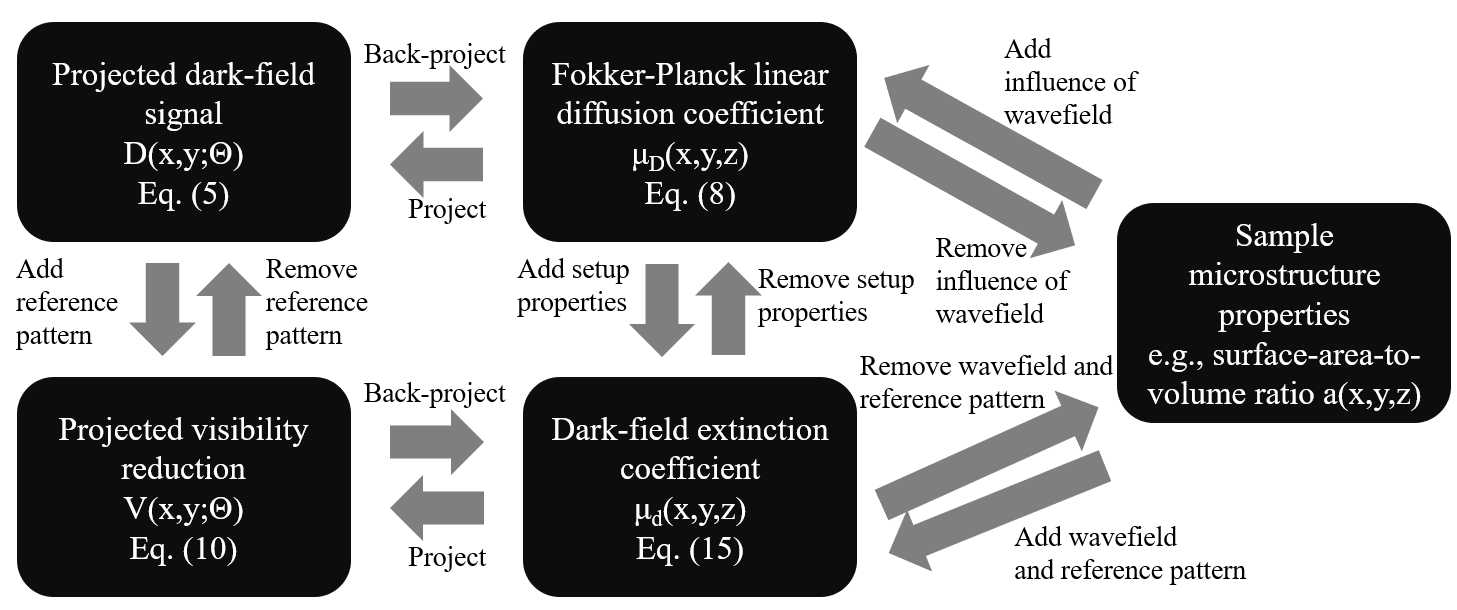}
    \caption{Correspondence between dark-field signal in the context of propagation-based imaging and visibility reduction in imaging methods that make use of reference patterns\cite{Bech2010,Lynch2011}. This figure also indicates the relationship between dark-field signal/visibility reduction and sample microstructure properties, such as surface-area-to-volume ratio.}
    \label{fig:SAVR_from_df}
\end{figure}
\noindent \par
To summarize, this paper has presented a computed tomography extension of the phase and dark-field retrieval method of Ref.~\cite{Leatham2023}, based on the Fokker-Planck generalization of the transport-of-intensity equation, allowing for the reconstruction of the sample linear attenuation coefficient and Fokker-Planck linear diffusion coefficient using the experimental technique of propagation-based imaging. In particular, we introduced a new measure, $\mu_{D}$, to quantify how much a given voxel of the sample will diffuse the x-ray beam, independent of the experimental set-up used to capture this measurement. The method presented provides an improvement in spatial resolution for the reconstruction of the sample linear attenuation coefficient CT slices compared to TIE-based phase retrieval. Additionally, our method provides complementary sample information about unresolved microstructure through the Fokker-Planck linear diffusion coefficient, providing one of the first demonstrations of x-ray diffusive dark-field tomography without optics. The resulting map of the Fokker-Planck linear diffusion coefficient could potentially be used in the future to infer sample measurements, such as the surface-area-to-volume ratio of the spatially unresolved sample microstructure.
\section*{Funding}
Australian Government Research Training Program (RTP) Scholarship; ARC Future Fellowship FT180100374, ARC Discovery Project DP230101327.
\section*{Acknowledgments}
This research was undertaken on the Imaging and Medical Beamline at the Australian Synchrotron, part of ANSTO, under proposal 18642. In particular, we thank Dr. Daniel Hausermann, Dr. Chris Hall and Dr. Anton Maksimenko for their help with the experiment. 
\section*{Disclosures}
The authors declare no conflicts of interest. 
\section*{Data availability}
Data underlying the results presented in this paper are not publicly available at this time but may be obtained from the authors upon reasonable request.
%%%%%%%%%%%%%%%%%%%%%%% References %%%%%%%%%%%%%%%%%%%%%%%%%

%%%%%%%%%% If using BibTeX:
\bibliography{references}

%%%%%%%%%% If preparing manually:
% \newcommand{\enquote}[1]{``#1''}

% \bibitem{Zhang:14}
% Y.~Zhang, S.~Qiao, L.~Sun, Q.~W. Shi, W.~Huang, L.~Li, and Z.~Yang,
%   \enquote{Photoinduced active terahertz metamaterials with nanostructured
%   vanadium dioxide film deposited by sol-gel method,}
%   {\protect\JournalTitle{Optics Express}} \textbf{22}, 11070--11078 (2014).

% \bibitem{Optica}
% {Optica}, \enquote{{Optica Publishing Group},}
%   \url{http://www.opg.optica.org}.

% \bibitem{FORSTER2007}
% P.~Forster, V.~Ramaswamy, P.~Artaxo, T.~Bernsten, R.~Betts, D.~Fahey,
%   J.~Haywood, J.~Lean, D.~Lowe, G.~Myhre, J.~Nganga, R.~Prinn, G.~Raga,
%   M.~Schulz, and R.~V. Dorland, \enquote{Changes in atmospheric consituents and
%   in radiative forcing,} in \enquote{Climate Change 2007: The Physical Science
%   Basis. Contribution of Working Group 1 to the Fourth assesment report of
%   Intergovernmental Panel on Climate Change,}  S.~Solomon, D.~Qin, M.~Manning,
%   Z.~Chen, M.~Marquis, K.~B. Averyt, M.~Tignor, and H.~L. Miler, eds.
%   (Cambridge University Press, 2007).

% \end{thebibliography}
\end{document}